\documentclass[11pt, a4paper,fleqn]{article}
\usepackage[latin1]{inputenc}
\usepackage{hyperref,bookmark}
\usepackage{amssymb}
\usepackage{amstext}
\usepackage{graphics}
\usepackage{graphicx}
\usepackage{authblk}
\usepackage{amsmath}
\usepackage{multirow}
\usepackage{xfrac}
\usepackage{ulem}
\usepackage[dvipsnames,svgnames]{xcolor}
\usepackage{listings} 

\usepackage{fdsymbol} 
\usepackage{mathtools,tikz,caption}
\DeclareRobustCommand\sampleline[1]{%
  \tikz\draw[#1] (0,0) (0,\the\dimexpr\fontdimen22\textfont2\relax)
  -- (2em,\the\dimexpr\fontdimen22\textfont2\relax);%
}

\usepackage{contour}

\contourlength{0.8pt}

\newcommand{\bm}[1]{{\boldsymbol {#1}}}

\include{defs}

\begin{document}

\title{Geometric approaches to assessing the numerical feasibility for conducting matching-adjusted indirect comparisons}

\author[1,2]{Ekkehard Glimm\thanks{Contact author: ekkehard.glimm@novartis.com}}
\author[1]{Lillian Yau}

\affil[1]{Novartis Pharma AG, Basel, Switzerland}
\affil[2]{University of Magdeburg, Magdeburg, Germany}

\maketitle

\begin{abstract}
We discuss how to handle matching-adjusted indirect comparison (MAIC) from a data analyst's perspective. We introduce several multivariate data analysis methods to assess the appropriateness of MAIC for a given data set. These methods focus on comparing the baseline variables used in the matching from a study that provides the summary statistics, or aggregated data (AD) and a study that provides individual patient level data (IPD). The methods identify situations when no numerical solutions are possible with the MAIC method. This helps to avoid misleading results being produced. Moreover, it has been observed that sometimes contradicting results are reported by two sets of MAIC analyses produced by two teams, each having their own IPD and applying MAIC using the AD published by the other team. We show that an intrinsic property of the MAIC estimated weights can be a contributing factor for this phenomenon.
\end{abstract}


\section{Introduction}\label{intro}

Since its introduction in 2010 \cite{Signorovitch2010}, the matching-adjusted indirect comparison (MAIC) method has generated much interest in the clinical trials community: see \cite{Bourdin2018, Levy2019, Signorovitch2015, Song2019, Strand2019} among others. With many health-care stakeholders, for example the National Institute for Health and Care Excellence (NICE) in the UK, or the Institute for Quality and Efficiency in Healthcare (IQWiG) in Germany, requiring supplemental comparisons beyond the pivotal clinical trial data, MAIC offers an opportunity for comparing new treatment options to available treatments that are not used in head-to-head comparisons in clinical trials. In addition, MAIC has also been applied as a network similar to the classic connected network meta-analysis \cite{Saure2020}. Furthermore, overviews of the method along with simulated treatment comparisons (STC), another indirect comparison method, have been provided by \cite{Ishak2015} and \cite{Phillippo2016}, where the latter also offers practical considerations for their implementations with extensive R code. These R code for estimating MAIC weights are now part of two R packages,  \texttt{maic} \cite{Young2021} and \texttt{MAIC} \cite{Bennett2021}.  More recently \cite{Cheng2020} examined the statistical properties and performance of MAIC in the causal estimand framework and provides mathematical formulations of its identification assumptions.

MAIC is a propensity score matching method. It leverages individual patient data (IPD) from one or more studies and adjusts their average patient characteristics to match to published results, or aggregated data (AD) from another study for which IPD are not available, but which otherwise is conducted in the same underlying target patient population. Each patient in each of the IPD studies is assigned a weight, measuring the patient's propensity of being in the AD study. If IPD and AD share a common treatment, the MAIC is called anchored, otherwise it is called unanchored.

The first and foremost assumption of MAIC is that IPD and AD are from a common underlying patient population and hence in principle comparable. Clinical input is crucial in understanding whether two studies can be compared using the method. IPD studies must be checked against the AD study in terms of study design, targeted patient population, inclusion and exclusion criteria, etc. to ensure the common population assumption of the method can be reasonably justified.

Once the clinical aspects of the studies are assessed and the common population assumption is considered to be satisfied, the data analyst will need to perform the next step of numerical examination of the data before proceeding to implementing the method. This is necessary because even if the inclusion and exclusion criteria of the different data sources are similar, it can still happen that due to shifts in regions or centers, the average patient characteristics are quite different within the admissible range. Our experience with MAIC is that so far there has been no reliable method to assess the numerical feasibility of implementing MAIC. In this paper, we present tools that help to conduct statistical checks of the data after the clinical comparability of the studies has been confirmed.

In particular, we present three methods: the first checks whether AD is within the range of IPD in order to have a numerically meaningful results. It is intuitive to see that in the one-dimensional space, i.e. when there is only one baseline variable to match, if AD is, say, 60 whereas the IPD ranges from a minimum of 20 to a maximum of 55, MAIC will not be able to match the IPD data onto the AD. We provide a method to check this in general for high dimensional data. The second method provides a visual check of where the AD is located in relation to the individual observations in the IPD in a multi-dimensional space; and the third is a test to see whether the same sampling mechanism underlies both studies such that MAIC is rendered unnecessary.

The paper is organized as follows. Section \ref{introData} presents an artificial data set to assist with illustrating the methods. Section \ref{maic} is a brief overview of MAIC, and argues why additional tools for examining the data are needed. In Section \ref{methods}, we discuss the three methods of checking the IPD against AD before implementing MAIC. Section \ref{examples} provides two examples with real clinical trial data. In Section \ref{discussion} we summarize our findings. The statistics software R is used throughout this paper. We have collated the code into a package which is available by request.

Without loss of generality, we assume there is one study that provides IPD, and one study that provides AD. When there are multiple IPD studies, the method can be applied to each study individually. It can also be applied after the IPD studies having been pooled in certain ways, such as following a propensity score matching, where the baseline variables are first weighted for each patient. These weighted variables are then used in the methods introduced below.

For ease of discussion the following notation is used. Let $\bar{\bm{x}}=(\bar{x}_1,\ldots, \bar{x}_p)'$ be the $p$-dimensional vector of AD, and $\bm Y=\left(\bm{y}_1 \ldots \bm{y}_n\right)$ with $\bm{y}_i=\left(y_{i1},\ldots, y_{ip}\right)'$ be the $p\times n$-matrix of IPD, where $n$ is the number of subjects in IPD study. Furthermore, $\bar{\bm{y}}=(\bar{y}_1,\ldots, \bar{y}_p)'$ is the $p$-dimensional vector of IPD summary statistics that is the counterpart of $\bar{\bm{x}}$. The response variable from AD is denoted $\bar{r}_{AD}$, and for each individual in the IPD, $r_{i}$. In our discussion we only focus on assessing the similarity between $\bar{\bm{x}}$ and $\bar{\bm{y}}$.

\section{An illustrative example}\label{introData}
To assist with the discussion, we introduce a two-dimensional IPD set along with three different scenarios of AD means that are used for three separate matchings. We use two-dimensional data here purely for illustrative purposes, since the methods we discuss below are not really needed for such a low dimension. The real value of the methods is with higher-dimensional data where visualization techniques are no longer sufficient to understand the data constellations.

\begin{table}[ht!]
\caption{Summary statistics of the IPD and the means of three scenarios (A, B, and C) of AD\label{table0}}
\centering
\begin{tabular}{c|cccccc|ccc}
\hline
&\multicolumn{6}{c|}{IPD summary statistics} & \multicolumn{3}{c}{AD means}\\
Variable & Mean & Min & Q2 & Median & Q3 & Max & A & B & C\\ \hline
$y_1$ & 4.6 & 0.9 & 2.5 & 4.1 & 6.8 & 9.8 & 3.8 & 7.6 & 6.5 \\
$y_2$ & 5.0 & 0.8 & 3.2 & 4.9 & 6.6 & 9.2 & 5.4 & 2.5 & 7.7 \\ \hline
\end{tabular}
\end{table}

\begin{figure}[ht!]
  \centering
  \includegraphics[width=1\textwidth]{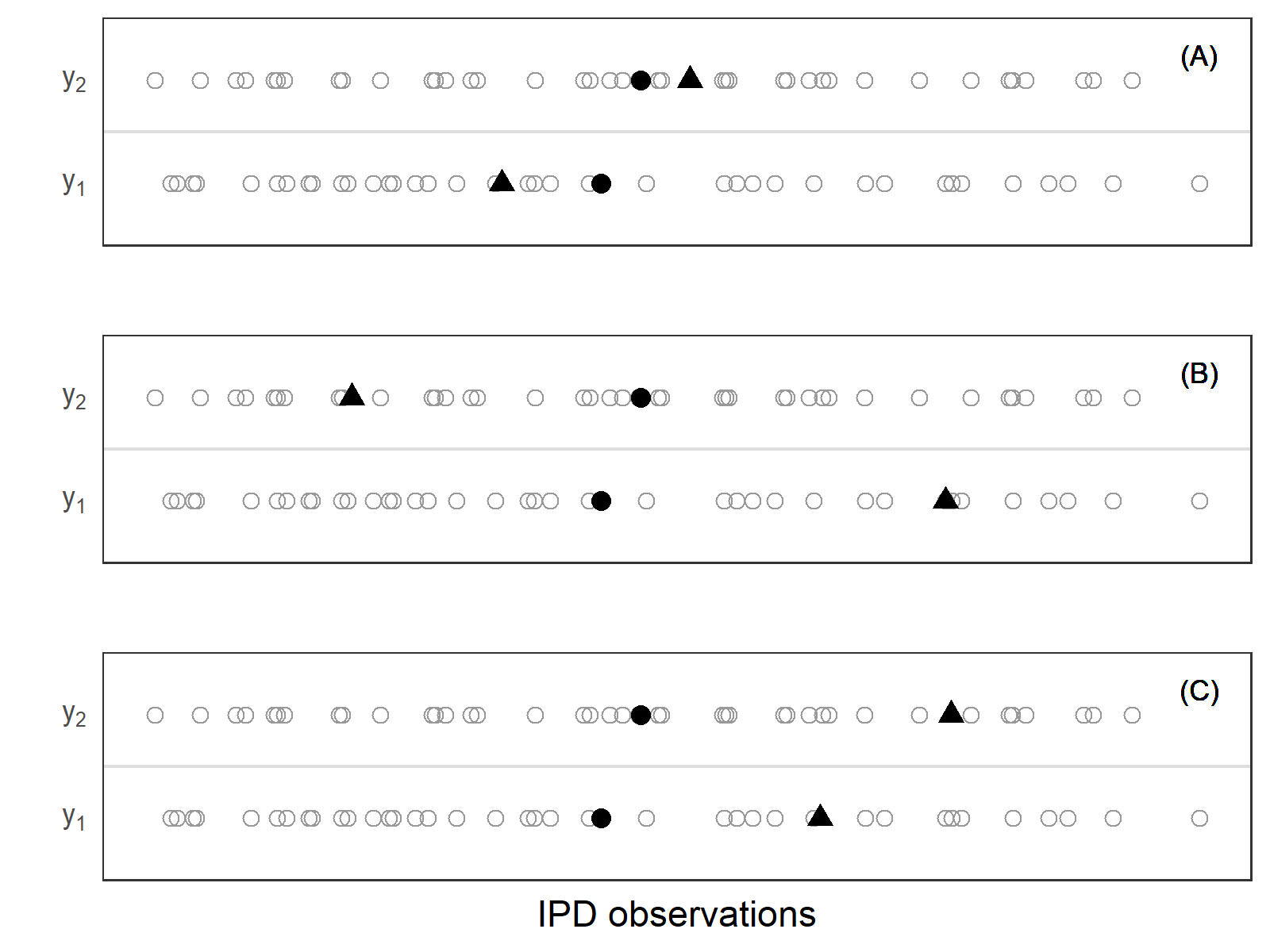}
  \centering \footnotesize{$\circ$ IPD $\bm{y}_i=(y_{i1}, y_{i2})'$ $\quad\quad$ \textbullet\ IPD center ($\bar{\bm{y}}$) $\quad\quad \smallblacktriangleup$ AD ($\bar{\bm{x}}$)}
 \caption{The introductory IPD and the AD summary statistics.}\label{p0}
\end{figure}

The summary statistics of the IPD and the AD are given in Table \ref{table0}. The marginal one-dimensional dot-plots are in Figure \ref{p0}, where Panels A, B, and C present the same IPD variables $y_1$ and $y_2$. In the figure the circles represent the individual observations, the black dots the IPD means. The three different AD scenarios are represented by the black triangles in the corresponding panels.

Summary statistics and dot-plots seem to indicate that all three scenarios of AD are well within the ranges of IPD variables. In Figure \ref{p0}(A)(Scenario A), AD is close to the IPD mean, especially in the case of $y_2$. In Figure \ref{p0}(B) (Scenario B), AD is further away from the IPD mean. In Figure \ref{p0}(C) (Scenario C), AD is on the upper end of the IPD ranges for both $y_1$ and $y_2$, but otherwise does not deviate from the IPD mean more than in Scenario B. Upon seeing these, one may hastily conclude that although AD does not coincide with the IPD mean exactly, there is no reason for alarm in any of the three scenarios.

\begin{figure}[ht!]
  \centering
  \includegraphics[width=1\textwidth]{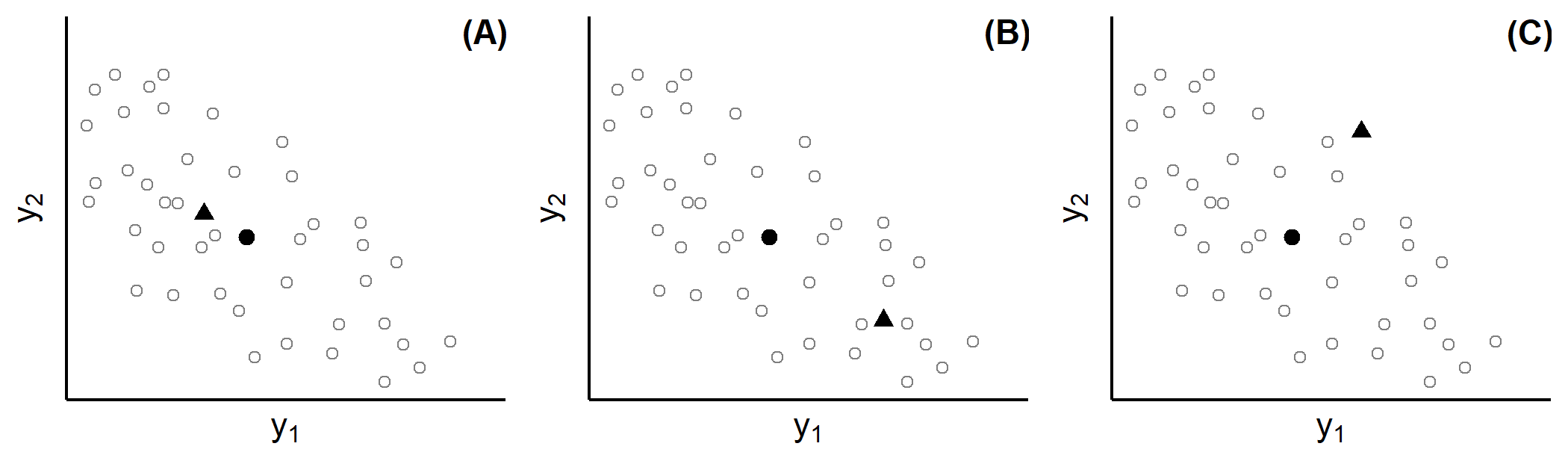}
  \centering \footnotesize{$\circ$ IPD $\bm{y}_i=(y_{i1}, y_{i2})'$ $\quad\quad$ \textbullet\ IPD center ($\bar{\bm{y}}$) $\quad\quad \smallblacktriangleup$ AD ($\bar{\bm{x}}$)}
 \caption{A two-dimensional IPD data and its corresponding AD summary statistics.}\label{p1}
\end{figure}

Since the data contain only two variables, the two-dimensional scatter plot in Figure \ref{p1} reveals the whole picture here. Although AD in Scenario C is closer to the IPD $\bm{\bar{y}}$ in terms of Euclidean distance than the AD in Scenario B, it is in fact situated outside the IPD cloud; whereas the AD in Scenario B is simply further away from $\bm{\bar{y}}$ but within the IPD cloud. Hence, no reweighing of IPD observations in Figure \ref{p0}(C) can exactly reproduce the AD $\bar{\bm x}$, whereas in Figure \ref{p0}(B), this is possible.

For data with three or more variables, looking at a two-dimensional scatter plot is like looking at the marginal one-dimensional dot-plots for a two-dimensional data set. Hence, for higher dimensional data, marginal two- or three-dimensional scatter plots will not necessarily reveal special features in the data. We need additional tools to help us examining any abnormalities in the data.

\section{An overview of MAIC}\label{maic}

MAIC assigns each patient $i,\ i = 1, ..., n$ in the IPD set a weight $w_i$ such that the summary statistics $\bar{\bm{y}}$ of the IPD matches the published AD summary statistics $\bar{\bm{x}}$. The weights $w_i$ are defined as
\begin{equation}\label{eq1}
w_i = \text{exp}(\bm{y}'_i\bm{\beta}),
\end{equation}
where $\bm{y}_i$ is a vector of baseline characteristics (or functions of them) to be matched. For example, if age is a given baseline covariate in the IPD, and average age is part of the vector $\bar{\bm{x}}$, then age is made one of the components of $\bm{y}_i$. If the variance of age is also given in the AD, then the squared residuals $\left(\sqrt{\frac{n}{n-1}}\left(y_{i, age}-\bar{y}_{age}\right)\right)^2$ can also be included as another component of $\bm{y}_i$.

MAIC estimates $\bm{\beta}$ by $\bm{\tilde{\beta}}$ so that the population moments based on the IPD study are equal to the published AD study sample moments. Hence $\bm{\tilde{\beta}}$ is the solution of
\begin{equation}\label{eq2}
\frac{\sum_{i=1}^{n}\bm{y_i}\text{exp}(\bm{y_i}'\bm{\tilde{\beta}})}
{\sum_{i=1}^{n}\text{exp}(\bm{y_i}'\bm{\tilde{\beta}})}=\bar{\bm{x}}.
\end{equation}

In their original publication of the MAIC method, the authors observed that equation (\ref{eq2}) constitutes a convex root finding problem and that hence, the solution for $\bm{\tilde{\beta}}$ is unique (up to a constant scalar) if there is such a solution \cite{Signorovitch2010}. However, in general, there is no guarantee that a solution exists.

If a solution exists, the estimated weight $\tilde{w}_i$ for each patient in IPD can be calculated from Equation (\ref{eq1}) by plugging in $\bm{\tilde{\beta}}$ in place of $\bm{\beta}$.

As with the propensity score matching method, an effective sample size (ESS) is given by
$$
\frac{\left(\sum_i\tilde{w}_i\right)^2}{\sum_i\tilde{w}_i^2}
$$
and is often used as a measure to retrospectively gauge the ``matchingness'' of IPD and AD data once the MAIC has been conducted. When ESS is close to the original sample size $n$, it is considered a good match. However, ESS is in fact only a measure of the variability of $\tilde{w}_i$'s. Higher ESS means less variability between $\tilde{w}_i$'s, which in turn can be due to a good fit between IPD and AD, but can also simply mean that important variables have not been included in the matching. Eliminating baseline characteristics from matching never decreases ESS.

Ultimately, we wish to compare the treatment effects between the IPD study and the AD study. To this end, $r_i$, the treatment outcome for patient $i$ in the IPD set, is multiplied by $\tilde{w}_i$. The weighted outcome can then be used in hypothesis testing or modeling as planned. For example, if a two-sample $t$-test is to be conducted, we would first calculate the MAIC adjusted mean treatment outcome in the IPD, i.e.
$$
\tilde{r}_{IPD}=\frac{\sum_{i=1}^{n}r_i\tilde{w}_i}{\sum_{i=1}^{n}\tilde{w}_i}.
$$
The re-weighted mean $\tilde{r}_{IPD}$ along with the ESS can then be used to test against $\bar{r}_{AD}$ as observed in the AD study.

\subsection{Implementing MAIC}\label{impMAIC}
\begin{figure}[ht!]
  \centering
  \includegraphics[width=1\textwidth]{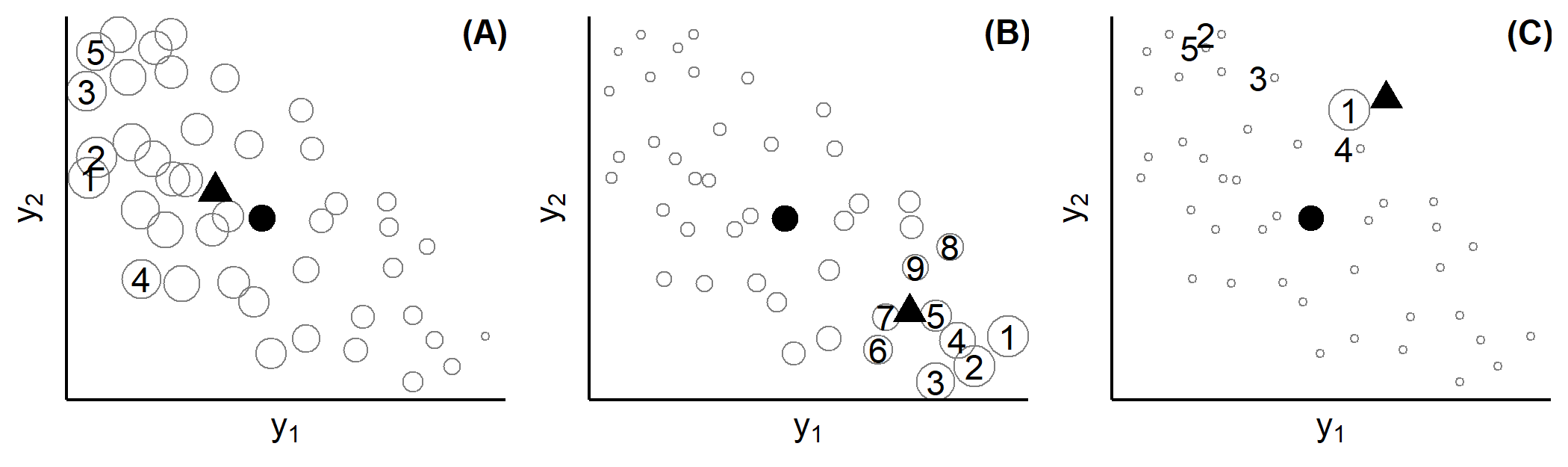}
  \centering \footnotesize{$\circ$ IPD $\bm{y}_i=(y_{i1}, y_{i2})'$ $\quad\quad$ \textbullet\ IPD center ($\bar{\bm{y}}$) $\quad\quad \smallblacktriangleup$ AD ($\bar{\bm{x}}$)}
 \caption{MAIC weights of the two-dimensional IPD data: larger circles indicating higher weight, with the largest few further labeled.}\label{p1wts}
\end{figure}

Since there is no closed form solution to Equation (\ref{eq2}), $\bm{\tilde{\beta}}$ are usually found through iterative algorithms such as the R function \texttt{optim()} \cite{Phillippo2016}, which we use here to apply MAIC to the data introduced in Section \ref{introData}. Table \ref{table0a} presents the MAIC weighted means for $y_1$ and $y_2$ of the IPD set together with the three AD's. The scatter plots of Figure \ref{p1} are reproduced in Figure \ref{p1wts} with the size of the circles indicating the estimated re-scaled MAIC weights, which are constrained to sum up to $n$, the total number of subjects in the IPD study. The observations with largest weights are further labeled in each panel with 1 being the patient receiving the highest weight. The size of the circles can only be compared within each panel and not between panels.

\begin{table}[ht!]
\caption{IPD MAIC-adjusted means and the three scenarios of AD\label{table0a}}
\centering
\begin{tabular}{c|ccc|ccc|ccc}
\hline
&\multicolumn{3}{c|}{Scenario A} & \multicolumn{3}{c|}{Scenario B} & \multicolumn{3}{c}{Scenario C}\\
Variable & \small{ESS/n} & IPD* & AD & \small{ESS/n} & IPD* & AD & \small{ESS/n} & IPD* & AD\\ \hline
$y_1$ & \multirow{2}{*}{88\%} & 3.8 & 3.8 & \multirow{2}{*}{31\%} & 7.6 & 7.6 & \multirow{2}{*}{2\%} & {\bf 5.7} & 6.5 \\
$y_2$ & & 5.4 & 5.4 & & 2.5 & 2.5 & & {\bf 7.4} & 7.7 \\ \hline
\multicolumn{10}{l}{\small{* IPD MAIC means}}\\
\end{tabular}
\end{table}

In Scenario A since $\bm{\bar{x}}$ is very close to $\bm{\bar{y}}$, we expect a good match between the IPD and the AD. This can be seen from the fairly uniform weights assigned to most of the subjects (Figure \ref{p1wts}(A). The ESS in this case is 88\% of $n$. In Scenario B $\bm{\bar{x}}$ is further away from $\bm{\bar{y}}$; therefore, only the cluster of subjects close to $\bm{\bar{x}}$ receive higher weights (Figure \ref{p1wts}(B)), and will contribute to the comparison of treatment outcomes of $\tilde{r}_{IPD}$ vs. $\bar{r}_{AD}$ more than the other subjects. The ESS is 31\% of $n$.

In Scenario C (Figure \ref{p1wts}(C)), since $\bm{\bar{x}}$ is outside of the IPD cloud, a solution to Equation (\ref{eq1}) does not exist. However, the search algorithm of \texttt{optim()} simply runs through the iterations, finds some result and returns it. The solution returned as such is arbitrary and not guaranteed to give an optimum. If the algorithm is allowed to run more iterations, a different result will often be returned. In this example, only one subject (the point labeled with 1 in Figure \ref{p1wts}(C)) is identified by the algorithm to receive a weight clearly larger than zero, resulting in an ESS of 2\% of $n$. What is more, as we can see from Table \ref{table0a} the MAIC IPD means do not equal the AD mean. In short, MAIC should not be applied to Scenario C, and it is important to first ascertain that the solution indeed exists before moving ahead with searching for a solution.

\subsection{The monotonic increases of MAIC weights along the line of steepest ascent}\label{monoInc}
Counter-intuitively, as we can see from Figure \ref{p1wts}, in scenarios A and B the largest weights are not given to the subjects in the IPD who are closest to $\bm{\bar{x}}$. This, in fact, is a consequence of the weight function as defined in Equation (\ref{eq1}). In Appendix \ref{appb} we show that there is a line of the steepest ascent and the weight must increase monotonically along this line, which leads to the highest weights going to the edges of the IPD data cloud rather than to subjects closest to $\bm{\bar{x}}$. 

There have been reports of contradicting MAIC results when company A compares their IPD with published AD from company B, vs. that when company B compares their IPD with published AD from company A \cite{Phillippo2016}. This is in line with our own experience from attending presentations and discussing the method with other data analysts. The monotonicity property of MAIC weights can be a contributing factor to this phenomenon. In Supplemental Material \ref{aweights} we briefly review some ideas for estimating the weights in ways that allocate higher weights to observations close to $\bm{\bar{x}}$.

\section{Methods for checking numerical feasibility}\label{methods}
\subsection{Checking whether a solution exists with LP-solve}\label{existSoln}
\begin{figure}[ht!]
  \centering
  \includegraphics[width=1\textwidth]{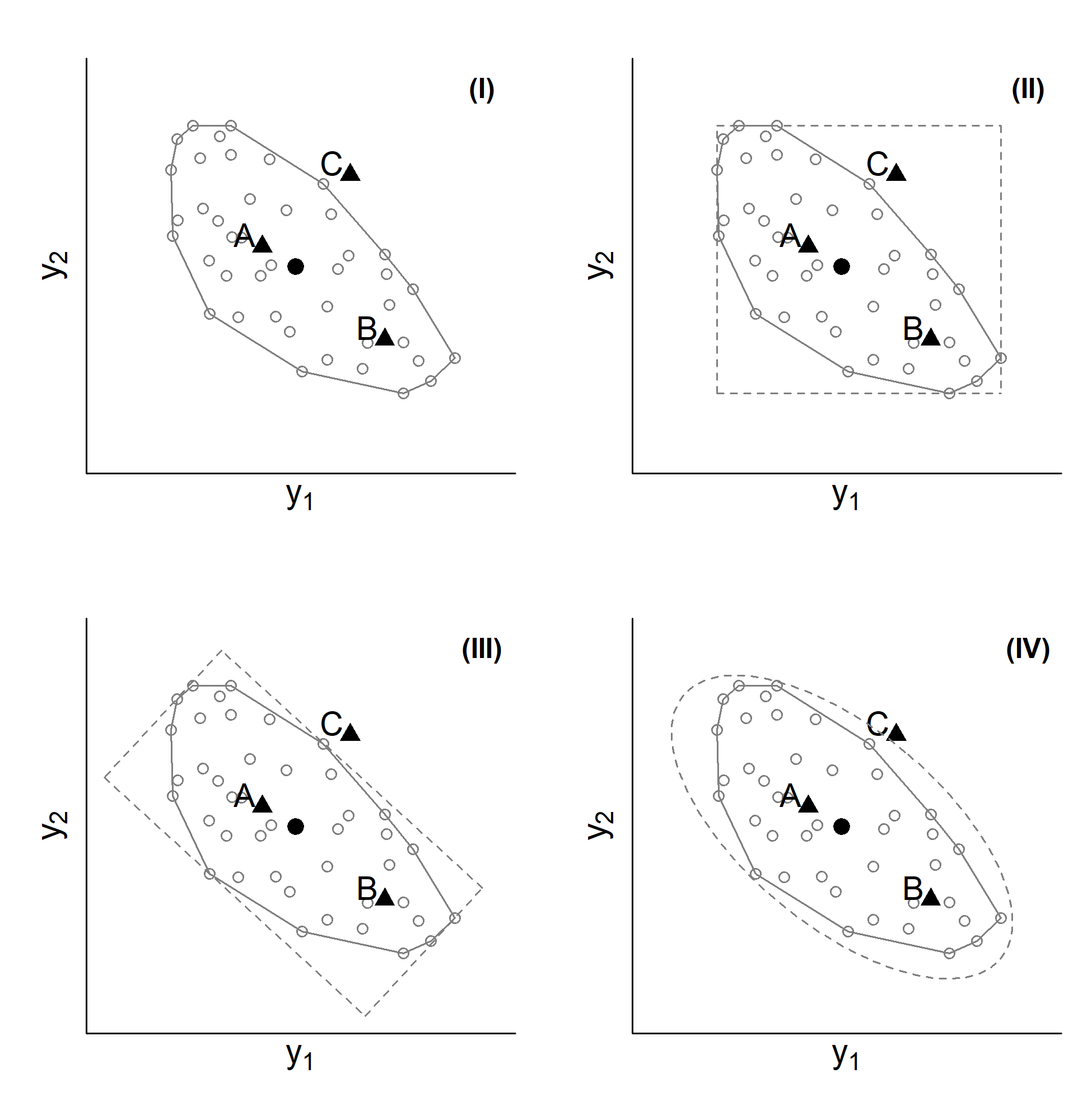}
  \centering
  \centering \footnotesize{$\circ$ IPD $\quad\quad$ \textbullet\ IPD center ($\bar{\bm{y}}$) $\quad\quad \smallblacktriangleup$ AD ($\bar{\bm{x}}$)}
 \caption{(I) A set of two-dimensional IPD points and its convex hull $H(\bm{Y})$; (II) an approximation of $H(\bm{Y})$ with observed ranges of IPD; (III) an approximation $H(\bm{Y})$ using principal components; (IV) an approximation of $H(\bm{Y})$ using Mahalanobis distance. The AD's corresponding to the three scenarios (A, B, and C) are labeled as such.}\label{p2ch}
\end{figure}

A \textit{convex hull} is the smallest convex shape that encloses all points in a set \cite{Lay1982}. In Figure \ref{p2ch}(I) the joined line segments enclosing the data cloud is the convex hull constructed from the two-dimensional IPD introduced in Section \ref{introData}. If $\bar{\bm{x}}$ is inside the enclosure, then we have numerically compatible data to apply MAIC method to.

More formally, Equation (\ref{eq2}) has a finite solution if and only if 
$\bar{\bm{x}}$ is an element of the set $H(\bm{Y})=\left\{\bm{z}: \bm{z}=\bm{Y}\bm{w}, 0\leq w_i\leq 1, \sum_{i=1}^n w_i=1\right\}$, $\bm{w}=\left(w_1, \ldots, w_n\right)'$. If $\bar{\bm{x}}$ is on a boundary of $H(\bm{Y})$, it can be represented by a subset of the $\bm{y}_i$'s, those that are sufficient to describe the corresponding bounding hyperplane. Hence, some patients---those not in the subset---will get a weight of $0$, bringing down the ESS.

The convex hull of any set of points in a $p$-dimensional space is a polygon (see Figure \ref{p2ch}(I) for an example with $p=2$). Finding this polygon is a famous problem in computational geometry that requires laborious iterative search.

In our application, however, it is sufficient to know whether $\bar{\bm{x}}$ is inside of $H(\bm{Y})$. This, in fact, is a much simpler and speedier task with the use of LP-solve. LP-solve is a generic term for any implementation of the simplex algorithm for optimizing linear functions under linear constraints. The method was originally developed by G. B. Dantzig in 1947 \cite{Dantzig1947}. In Appendix \ref{appa} we show how to set up LP-solve for this particular task. In our accompanying R package we implement the method using the function \texttt{lp()} in the R package \texttt{lpSolve} \cite{Berkelaar2020}.

The procedure as we set it up takes $\bar{\bm{x}}$ and $\bm Y$ as input, and checks whether $\bar{\bm{x}}$ is an element of $H(\bm{Y})$. If it is, \texttt{lp()} returns an exit code of 0; otherwise an exit code 2 is returned.

Applying the method, the procedure confirms that $\bar{\bm{x}}$ in Figure \ref{p1}(A) and \ref{p1}(B) are both inside $H(\bm{Y})$. For Figure \ref{p1}(C) however it shows that $\bar{\bm{x}}$ is outside $H(\bm{Y})$. Knowing this, we could proceed with MAIC for the former two cases, but would not for data in Figure \ref{p1}(C).

\subsection{Checking where AD is with principal component analysis}\label{pca}

Besides numerically checking whether a solution exists, visual assessment can often provide additional information. For example, once it is ascertained that $\bar{\bm{x}}$ is within $H(\bm{Y})$, naturally we would like to know how close it is to $\bar{\bm{y}}$. We introduce a graphical approach to aid us with this assessment using principal component analysis (PCA).

PCA is a commonly used dimensionality reduction technique. It is an orthogonal linear transformation of a set of data to a new coordinate system such that the largest variation of the data is along the first coordinate (called the first principal component (PC)), the second largest variation on the second coordinate, and so on \cite{Manly1994}.

In our application, we use PCA to approximate the convex hull $H(\bm{Y})$ with a hyper-rectangle whose bounding hyper-planes are parallel to the PCs (Figure \ref{p2ch} (III)). As illustrated in the figure, representing the IPD in terms of PCs provides a tighter box around the IPD than the original coordinates (Figure \ref{p2ch} (II)). In many cases, it is a good approximation of the minimum-volume bounding box \cite{Dimitrov2012}. Hence, it can also detect cases where AD is outside of $H(\bm{Y})$ more often than just checking the ranges of the original IPD variables.

However, the main purpose of PCA in this application is to visualize how $\bm{\bar x}$ is situated relative to $\bm{\bar y}$ in IPD PC coordinates; hence, we recommend only proceed with the PCA after having checked whether $\bar{\bm{x}}$ is inside $H(\bm{Y})$ with LP-solve.

To implement the method, we first compute PCs for the IPD. For numerical reasons, as it is often the practice of PCA, it is recommended to first standardize the IPD and the AD variables with respect to the means and standard deviations of IPD data. If there are $p$ variables in IPD, there will be $p$ PCs (except for pathological cases where for example $n<p$, or one of the $p$ variables is a linear combination of others). The next step is to project AD onto each of the $p$ IPD PC coordinates using the matrix of the variable loadings produced by PCA. We can then check where $\bar{\bm{x}}$ is in relation to $\bar{\bm{y}}$ and the rest of IPD in PC coordinates. Although having $\bar{\bm{x}}$ within the range of all the PCs does not always imply that AD is inside the IPD cloud, the converse is always true. If $\bar{\bm{x}}$ is outside the range of any PCs, then we know for sure that AD is outside IPD.

\begin{figure}[ht!]
  \centering
  \includegraphics[width=1\textwidth]{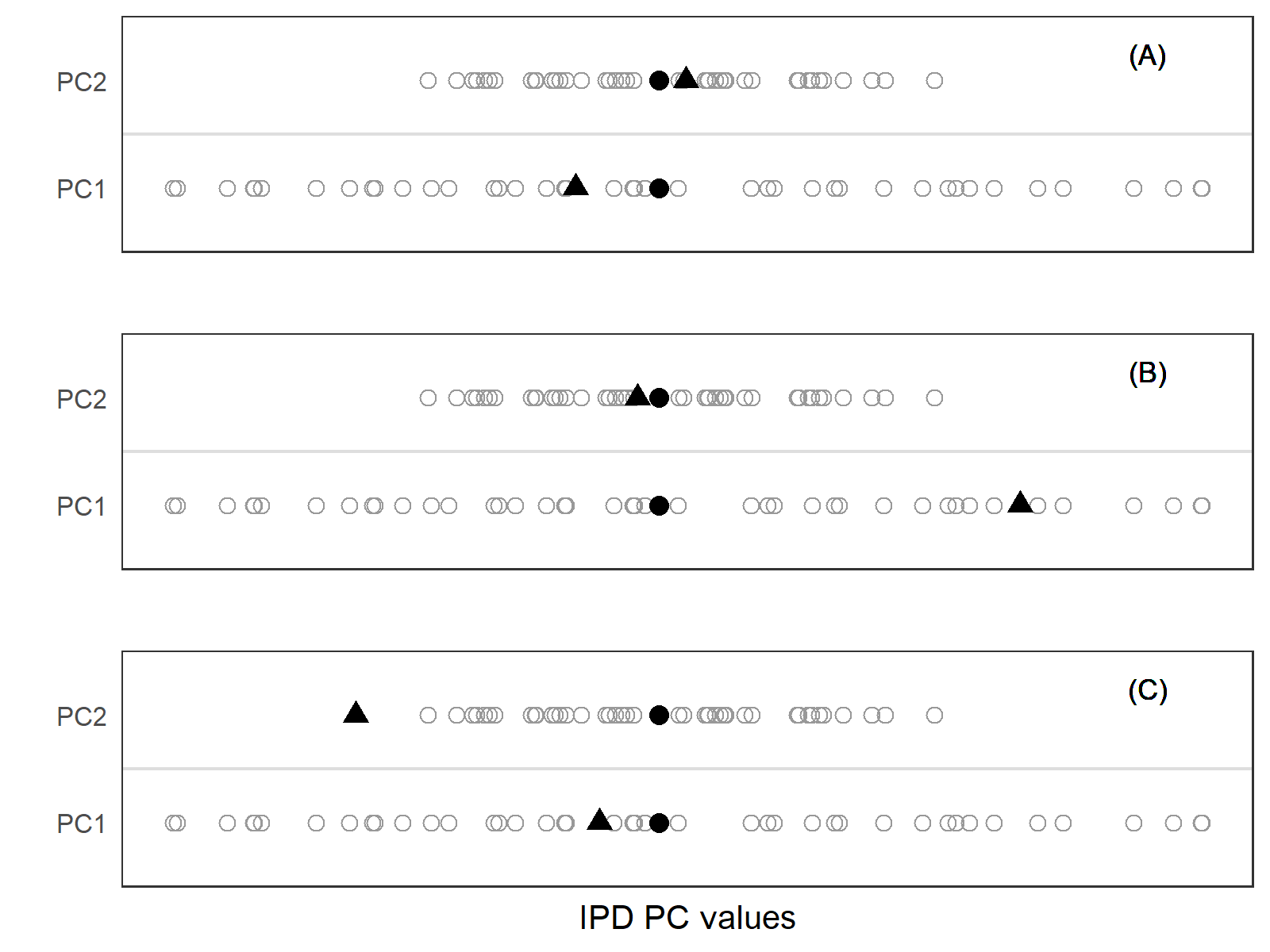}
  \centering
  \centering \footnotesize{$\circ$ IPD $\quad\quad$ \textbullet\ IPD center ($\bar{\bm{y}}$) $\quad\quad \smallblacktriangleup$ AD ($\bar{\bm{x}}$)}
 \caption{IPD PC with AD in PC coordinates}\label{p1pc}
\end{figure}

Figure \ref{p1pc} presents the results after implementing the method for the three scenarios from Section \ref{introData}. In Figure \ref{p1pc}(A), since $\bar{\bm{x}}$ is very close to $\bar{\bm{y}}$ to begin with, in the PC coordinates, it is no surprise that AD is well within the ranges of all PCs and very close to the center. In Figure \ref{p1pc}(B), AD is further away from the IPD center, but still within the ranges. In Figure \ref{p1pc}(C), however, AD is clearly outside the range of PC2, indicating that AD is outside of $H(\bm Y)$.

\subsection{Checking whether MAIC is at all necessary using Mahalanobis distance}

When all matching variables are from a multivariate normal distribution, $H(\bm{Y})$ can be approximated by the contour of a corresponding multi-dimensional ellipsoid. This ellipsoid is the set of all points with a Mahalanobis distance smaller or equal to the largest Mahalanobis distance in the IPD set, that is
$$
\{\bm{y}: (\bm{y}-\bar{\bm{y}})'\bm\Sigma^{-1}(\bm{y}-\bar{\bm{y}})\ \le \max_i (\bm{y}_i-\bar{\bm{y}})'\bm\Sigma^{-1}(\bm{y}_i-\bar{\bm{y}})\},
$$
where $\bm{\Sigma}$ is the covariance matrix of the multivariate normal distribution. In a two-dimensional space, the ellipsoid is an ellipse (Figure \ref{p2ch} (IV)).
If the Mahalanobis distance of $\bar{\bm{x}}$ from $\bar{\bm{y}}$ is larger than the distances of all $\bm y_i$ from $\bar{\bm{y}}$, then $\bar{\bm{x}}$ is outside of $H(\bm{Y})$. The issue with this approach is that the Mahalanobis distance is designed for elliptically contoured distributions, of which the multivariate normal distribution is by far the most important one. In practice, MAIC is often used on baseline characteristics such as age, sex, disease status, etc. Many of these variables are binary or categorical, and a multivariate normal assumption is unrealistic.

On the other hand, the Mahalanobis distance can be useful for other purposes. For example, once we know that $\bar{\bm{x}} \in H(\bm{Y})$ we can apply the method for testing whether matching IPD to AD is at all necessary.

It is known that the Mahalanobis distance is closely related to both discriminant analysis and Hotelling's $T^2$-test. The latter is a test of the mean of a multivariate normal distribution. In this context, the test statistic of Hotelling's $T^2$-test,
\begin{equation}\label{eq3}
T^2=n\left(\bar{\bm{y}}-\bar{\bm{x}}\right)'\hat{\bm\Sigma}^{-1}\left(\bar{\bm{y}}-\bar{\bm{x}}\right)
\end{equation}
is proportional to the Mahalanobis distance between the IPD data points and the AD. Here, $\hat{\bm\Sigma}$ is the covariance matrix estimated from the IPD data. If the data is multivariate normal, then $\frac{n-p}{pn-p}T^2$ has an $F$-distribution with $(p, n-p)$ degrees of freedom (assuming that $\bar{\bm{x}}$ is fixed). Alternatively, if the sample size $n_{AD}$ of the AD data is known, we can standardize the Mahalanobis distance via
\begin{equation}\label{eq4}
T^2_{AD}=\frac{n\cdot n_{AD}}{n+n_{AD}}\cdot\left(\bar{\bm{y}}-\bar{\bm{x}}\right)'\hat{\bm\Sigma}^{-1}\left(\bar{\bm{y}}-\bar{\bm{x}}\right).
\end{equation}
In this case, $\frac{n-p}{pn-p}\cdot T^2_{AD}$ has an $F$-distribution with $(p, n-p)$ degrees of freedom if the data is multivariate normal. This is different from the distribution of the ``usual" two-sample-$T^2$-test, because we estimate $\bm\Sigma$ only from the IPD data, but still consider $\bar{\bm{x}}$ as subject to random variation.

We can hence use the Mahalobis distance and Hotelling's $T^2$ to test if the IPD and the AD may have been generated by replicating the same sampling mechanism in the common target population. In Equation (\ref{eq3}), the AD data is treated as given, and Hotelling's $T^2$ tests if $\bm \mu_{IPD}=\bar{\bm{x}}$ is plausible. In Equation (\ref{eq4}), the test of $T^2_{AD}$ investigates whether both IPD and AD might be sampled from the same underlying distribution with a common multivariate mean. We note in passing that since for AD, we only have $\bar{\bm{x}}$, the latter test can only operate under the (untestable) assumption that the covariance matrices of AD and IPD are the same.

If $n$ is large in relation to $p$, then due to asymptotic theory $\bar{\bm{y}}$ will be approximately normally distributed. Hence, for testing we do not need a multi-normal distribution assumption for the entire IPD set. When $\bar{\bm y}$ cannot be approximated well with a multi-normal distribution, we could still use $T^2$ or $T^2_{AD}$ in a resampling test.

A large $p$-value of the test would indicate that matching is not necessary and simply pooling the data is acceptable. Applying the method to Scenarios A and B in our introductory example, the Hotelling's $T^2$-tests give $p$-values of $> 0.1$ and $< 0.0001$, respectively, when using Equation \ref{eq3}. This suggests that for Scenario A MAIC is not necessary, and a direct comparison of the outcome variables between the two studies is appropriate. The $p$-value of $< 0.0001$ for Scenario B indicates that matching of baseline variables is necessary for a fair comparison between the two studies.

\section{Examples}\label{examples}
\subsection{Example 1}\label{ex1}
In a recent randomized clinical study, we compared the efficacy of a new treatment (T1) head-to-head with an active control (T2). Treatment T1 has demonstrated superiority over T2. While the clinical study was ongoing, another treatment, T3, became available. Since we were not able to directly compare T3 in the clinical study, we wish to conduct an indirect comparison using IPD from our study and the published AD on T3. An unanchored MAIC is performed due to the lack of a common comparator. In addition, we use only a subset of the data in order to mask the study for confidentiality.

After discussing with the clinicians, we identified the relevant baseline variables to match. These variables include patient demographics, disease history and characteristics, and prognostic factors relevant to the disease. The two studies are conducted in the same target patient population. They are of similar size, both with number of patients in the 100 to 200 range.

Although some of these baseline variables are continuous in nature, only ranges and medians are reported by the published AD. As a consequence, these continuous variables have to be dichotomized (e.g. above the median vs. below the median). Original categorical variables with $k$ categories are dummy-coded with $k-1$ binary variables. This eventually leads to a total of nine binary variables to be matched.

With nine binary variables we have 288 configurations of possible outcomes of the baseline variables. To observe all of them we would need at least 288 patients in the IPD study. With less than 288 of patients in our study, there is a possibility that the AD might not be situated within the IPD convex hull. Therefore, it is essential to first check whether MAIC can be applied at all to warrant a numerically meaningful result.

\begin{figure}[ht!]
  \centering
  \includegraphics[width=1\textwidth]{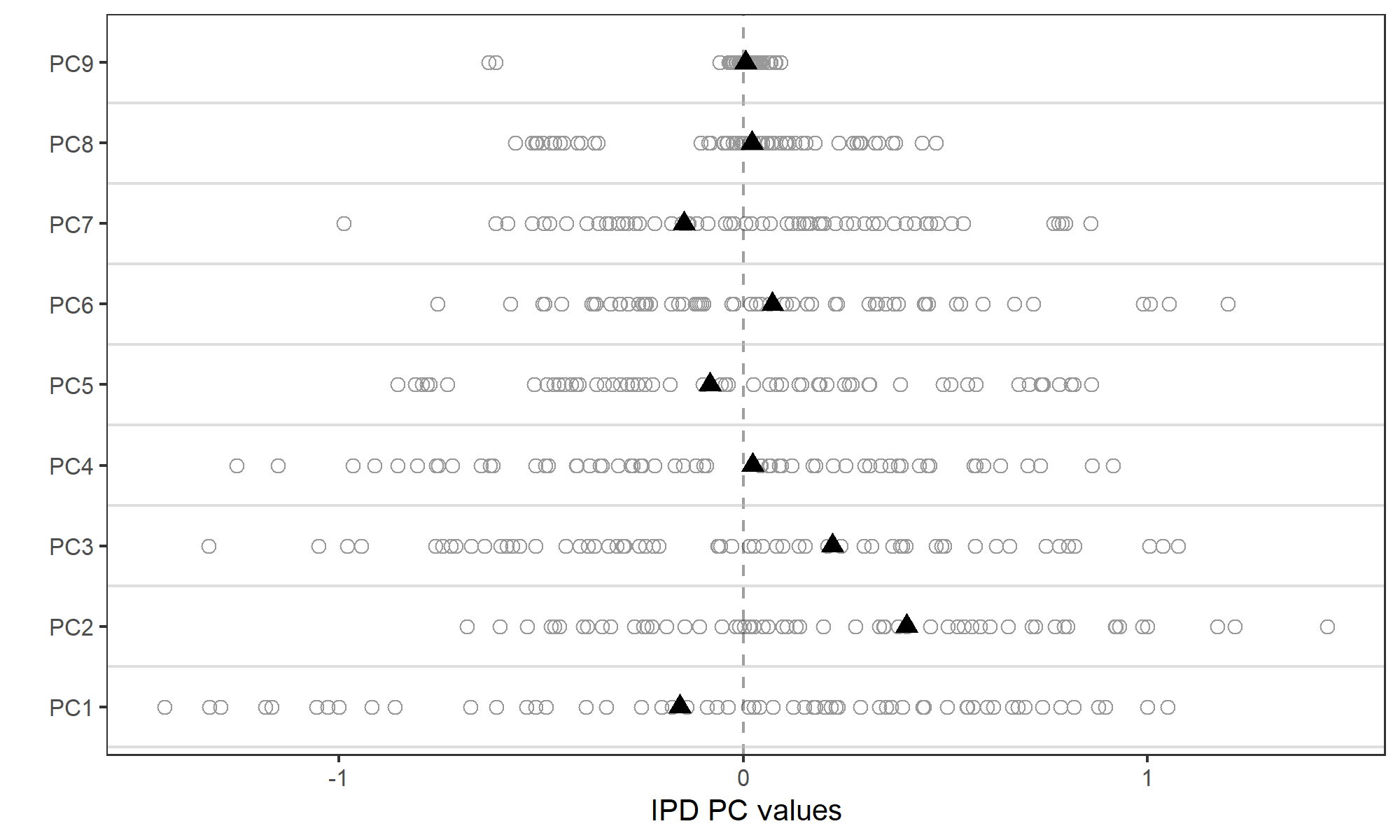}
  \centering
  \centering \footnotesize{$\circ$ IPD $\quad\quad$ \sampleline{dashed}\ IPD center ($\bar{\bm{y}}$) $\quad\quad \smallblacktriangleup$ AD ($\bar{\bm{x}}$)}
 \caption{IPD PC with AD in PC coordinates.}\label{pc9a}
\end{figure}

We first apply LP-solve to check whether AD is within the convex hull of the IPD. The procedure returns an exit code of 0, guaranteeing a unique solution to the MAIC weights. We next apply PCA on the IPD to have a visual check where the AD stands with regard to the IPD. The results are in Figure \ref{pc9a}.

Since there are nine derived baseline variables used in the matching, nine PCs are produced by the PCA. From the dot-plots of the PCs in Figure \ref{pc9a} we see that not only AD are well within the ranges of the IPD in PC coordinates, the AD are reasonably close to the centers (the vertical dashed line going through zero). 
As a result, MAIC can be carried out with no numerical complications.

Additionally applying Mahalanobis distance and Hotelling's $T^2$-test yields a $p$-value $< 0.0001$. Matching and adjusting the baseline covariates before comparing the treatment outcome is necessary.

\subsection{Example 2}\label{ex2}
In the same publication where we obtained the AD in Example 1, the authors also reported separately results for a subgroup of patients who exhibit a certain mutation. These results were not part of the AD in the previous example. Internally we identified patients with the same mutation in another randomized study where treatment T1 was tested. We would like to repeat the exercise and compare the treatment effect of T1 and T3 in this subgroup. Since one of the variables is not available in this IPD set, we have eight variables to match.

\begin{figure}[ht!]
  \centering
  \includegraphics[width=1\textwidth]{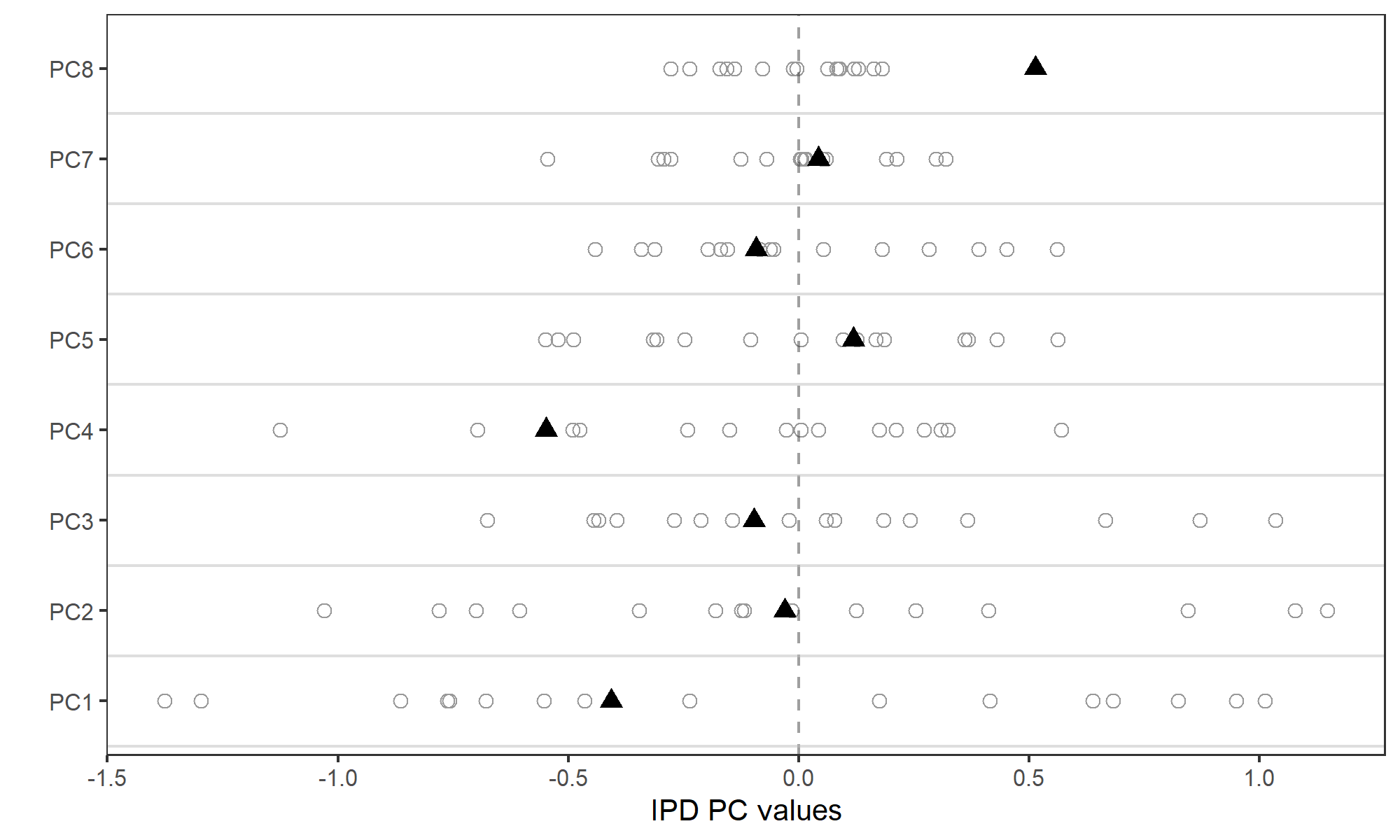}
  \centering
  \centering \footnotesize{$\circ$ IPD $\quad\quad$ \sampleline{dashed}\ IPD center ($\bar{\bm{y}}$) $\quad\quad \smallblacktriangleup$ AD ($\bar{\bm{x}}$)}
 \caption{IPD PC with AD in PC coordinates.}\label{pc8x}
\end{figure}

Applying LP-solve reveals that there is no solution to Equation (\ref{eq2}), and the PCA plot (Figure \ref{pc8x}) also shows that AD PC coordinate 8 is outside the range of PC8. Therefore, we cannot conduct MAIC in this subgroup.

\section{Discussion}\label{discussion}

We have illustrated how LP-solve can be used to check if IPD and AD are numerically compatible to give a solution using the MAIC method.
One may argue that \texttt{optim()} in R or other optimization functions also provide a convergence status when used to find MAIC weights.
However, these ``all-purpose" optimizers are designed to find optima of general functions. They neither check whether solutions to the optimization problem exist, nor indicate whether a global optimum, a local optimum or some other value is identified. Our experience shows that in case of a non-existing solution, different starting values and number of iterations can lead to different numerical results in addition to an unreliable convergence message.

In contrast, LP-solve always reports the correct solution in finite steps without any dependence on starting values, the given number of iterations or a threshold for declaring convergence. As an illustration of this point, when we ran \texttt{optim()} in Example 2 of Section \ref{ex2} the function eventually converged, if only the maximum number of iterations was set to a high enough value. However, comparing the re-weighted summary statistics of IPD to its target AD, while many of the variables converged toward AD (but not exactly equal to AD), some actually moved away from it.

On the other end of the spectrum, when IPD and AD data are already similar before performing MAIC, it may be unnecessary to apply MAIC. As discussed in Section \ref{maic}, the MAIC weights have the potentially undesirable property that they monotonously increase in certain directions of the multivariate space, which implies that the highest weights are not given to observations close to the AD data, but rather to some ``extreme" IPD observations (see Figure \ref{p1wts}).  If there are indications that IPD and AD might well have been obtained by the same sampling mechanism from a common population, then we should be less inclined to accept this warping of the multivariate distribution of IPD observations by MAIC weights. To assess the plausibility of a common sampling, we can use the Hotelling's $T^2$-test. It is important to emphasize that in this context the test is not used as a formal confirmatory decision procedure, but rather as an indicator of how close IPD and AD are, similar to a goodness-of-fit test in many statistical models.

We note in passing that LP-solve also opens up a way to come up with alternative weights for matching IPD to AD. As the objective function of LP-solve, we can use $\bm{c}'\bm{v}$ with $\bm{c}$ being a projection of any $n$-dimensional vector $\bm{z}$ onto the space orthogonal to the IPD $\bm{Y}$. By using different $\bm{z}$'s, we obtain many sets of eligible weights whose linear combinations are again eligible weights which exactly match IPD and AD if there is such an exact match. In this way, we can potentially produce weights that have certain properties, such as higher weights for IPD observations close to the AD. More detail is provided in Supplemental Material \ref{aweights}. A thorough investigation of alternative weights is, however, a topic for future research.

In conclusion, the suggested methods for checking numerical feasibility of MAIC are easy and straightforward to implement. We have simply re-purposed a few tried-and-true methods from the multivariate data analysis toolbox. We have developed an accompanying R package, in which the data only need to be in exactly the same format as necessary for fitting MAIC, so that no additional data processing is required. The running time for each method is next to null, since no iterative searches or simulations are required. We believe these methods can help data analysts to best utilize the strength of MAIC, and to avoid generating misleading results when the method should not be used.

\appendix
\section{Appendices}

\subsection{Implementing LP-solve}\label{appa}

We proceed as follows:
\begin{enumerate}
\item Set up some objective function $\bm{c}'\bm{v}$ to maximize w.r.t. $\bm{v}$. For checking for the existence of a solution, it does not matter what $\bm{c}$ is. If to be used for alternative weights, see Supplemental Material \ref{aweights}.
\item Set the constraints:
\begin{eqnarray*}
&&\bm{Y}\bm{v}=\bar{\bm{x}}\\
&&\bm{1}_n'\bm{v}=1 \mbox{ (the sum of components in $\bm{v}$ is $1$) }\\
&&v_i\geq 0 \mbox{ (all components $v_i$ of $\bm{v}$ are 0 or positive). }
\end{eqnarray*}
\end{enumerate}
Step 1 of this setup is not really relevant here, since we merely use LP-solve's simplex algorithm to check if the constraints have a solution. In contrast to finding the convex hull, this is performed very quickly and efficiently in a finite number of steps and the algorithm immediately returns whether there is a solution, which is all we need for this application.

\subsection{The monotonic increases of the MAIC weights along the line of the steepest ascent}\label{appb}

In MAIC, the weight for the $i^{th}$ subject in the IPD is given by $w_i\propto \exp(\bm{y}_i'{\bm \beta})$ subject to $\sum_{i=1}^n w_i \bm{y}_i = \bar{\bm{x}}$ to ensure uniqueness of $w_i$. Since $w_i\propto \exp(\bm{y}_i'{\bm \beta})$ is a monotonous function in all components $y_{ij}$ of $\bm{y}_i$, for $y_{ij} \rightarrow \infty$, $w_i$ will either go to $\infty$ or to $0$, depending on the sign of $\beta_j$ when $y_{ij'},\ j' \neq j$ remain the same for all $j'$.

In order to better understand the precise behavior of the weights, we can calculate in which direction they increase fastest in the $p$-dimensional space of $\bm \beta$ and $\bm{y}_i$'s. To simplify notation, without the loss of generality, we assume that $\bar{\bm{x}}=\bm{0}$. The steepest ascent for a given $\bm \beta$ will happen in the direction of the $\bm{y}$ which maximizes $\bm{y}'{\bm \beta}$ subject to a fixed length $\bm{y}'\bm{y}=c$ where $c$ is a constant that must be finite and larger $0$, but is otherwise irrelevant. Among all points $\bm{y}$ that are equally far away from $\bar{\bm{x}}=\bm{0}$, this maximum is the one getting the highest weight.

To see this, consider $\max\left(f(\bm{y})\right)$ where $f(\bm{y})=\frac{\bm{y}'\bm \beta}{\sqrt{\bm{y}'\bm{y}}}$ (the denominator is introduced to render the objective function independent of $c$). Calculating
$$
\frac{df(\bm{y})}{d y_j}=\frac{\beta_j\cdot \sqrt{\bm{y}'\bm{y}}-\left(\bm{y}'\bm{y}\right)^{-1/2}y_j\bm{y}'\bm\beta}{\bm{y}'\bm{y}},
$$
we see that $\frac{df(\bm{y})}{d y_j}=0$ if $y_j=\frac{\beta_j \bm{y}'\bm{y}}{\bm{y}' \bm \beta}$. Hence, the maximum of $\bm{y}'{\bm \beta}$ is achieved on any hypersphere $\bm{y}'\bm{y}=c$ when $\frac{y_j}{y_k}=\frac{\beta_j}{\beta_k}$ for all $1\leq j,k\leq p$.

If we switch the sign of $\bm{y}$ to $-\bm{y}$, we obtain the only other solution to $\frac{df(\bm{y})}{d y_j}=0$ which hence must be a minimum (to be precise, one of $\bm{y}$ and $-\bm{y}$ constitutes the minimum, the other the maximum, which is which is a matter of convention in notation and need not concern us here). Since $\bm{y}$ and $-\bm{y}$ are on the same axis, only on opposite sides of $\bar{\bm{x}}=\bm{0}$, this shows that the direction of $\bm{y}$ is also the one of steepest increase in the value of the weight, as for any other direction $\bm{y}^*$, the difference ${\bm{y}^*}'{\bm \beta}- (-\bm{y}^*)'{\bm \beta}$ must be smaller than $\bm{y}'{\bm \beta}- (-\bm{y})'{\bm \beta}$.

\section{Supplemental material}

\subsection{Alternative weights}\label{aweights}

The MAIC weights $w_i\propto \exp(\bm{y}_i'{\bm \beta})$ are not the only possible weights which fulfill the conditions $\bm{Y}\bm{w}=\bar{\bm{x}}$, $w_i\geq 0$ for all $i=1,\ldots,n$. Assuming without loss of generality that $\bar{\bm{x}}=\bm{0}$, the entire set $W$ of eligible weights is given by the intersection of the hyperspace orthogonal to $\bm{Y}$ and the positive orthant, i.e.
$$W=\left\{\bm{w}\in \bm{R}^n: \bm{w}=\bm{P}\bm{z} \mbox{ for any } \bm{z}\in \bm{R}^n, \bm{z}\neq \bm{0}\right\}\cap \left\{\bm{w}\in \bm{R}^n: w_i\geq 0 \forall i=1,\ldots, n\right\}$$ where $\bm{P}=\bm{I}_n-\bm{Y}'\left(\bm{Y}\bm{Y}'\right)^{-1}\bm{Y}$ is the projection matrix of the orthogonal projection. $W$ can be empty (except for the origin), but if it is not, it is a space containing infinitely many solutions. This means that another vector $\bm{w}$ of valid weights is found if for any arbitrary $\bm{z}\in \bm{R}^n, \bm{z}\neq \bm{0}$, all components $w_i\geq 0$. Since any projection $\bm{P}\bm{z}$ fulfills $\bm{Y}\bm{P}\bm{z}=\bm{0}$, we can again use LP-solve from Appendix \ref{appa} with $\bm{c}=\bm{P}\bm{z}$ to get a set of weights in $W$ (unless $W$ is empty) which are different from the MAIC weights.

Here, we give an example of how other weights than the MAIC weights might be produced: As shown in Appendix \ref{appb}, the MAIC weights display an undesirable monotonicity causing observations far away from $\bar{\bm{x}}$ to get a high weight. In order to avoid this, the following could be done:
\begin{enumerate}
\item Run LP-solve $n$ times with $\bm{c}$ being the $i$-th column of $\bm{P}$. The $i$-th column of $\bm{P}$ is the projection of the $i$-th axis onto the hyperspace orthogonal to $\bm{Y}$. This operation gives us $n$ sets of feasible weights $\bm{w}^{(k)}$, all producing $\bm{Y}\bm{w}^{(k)}=\bar{\bm{x}}$. However, many of the weights $w_i^{(k)}$ will be 0.
\item Since $\bm{Y}\bm{w}^{(k)}=\bar{\bm{x}}$ for all $k$, $\bm{Y}\bm{W}\bm{d}=\bar{\bm{x}}$, where $\bm{W}=\left(\bm{w}^{(1)}\cdots \bm{w}^{(n)}\right)$ and $\bm{d}=(d_1,\ldots, d_n)'$ with $\sum_k d_k=1$, also holds.
    Hence, $\bm{W}\bm{d}$ also is a feasible set of weights. To make sure that IPD observations $\bm{y}_k$ close to $\bar{\bm{x}}$ get higher weights, we choose $d_k$ inversely proportional to the Euclidean or Mahalanobis or some other distance of $\bm{y}_k$ from $\bar{\bm{x}}$.
    For example, with the Euclidean distance, we would use
    $d_k\propto \left(\left(\bm{y}_k-\bar{\bm{x}}\right)'\cdot \left(\bm{y}_k-\bar{\bm{x}}\right)\right)^{-1}$, standardized to sum up to 1. The ultimate weights would be $\bm{W}\bm{d}$. They are not monotonic with distance from $\bar{\bm{x}}$ (that would be an additional restriction rendering solvable IPD constellations insolvable), but they have a tendency to give high weights to IPD observations close to $\bar{\bm{x}}$.

\end{enumerate}

\subsection{R package}\label{appc}
The R package \texttt{maicChecks} is available upon request, as well as the R code for finding the alternative weight described in Supplemental Material \ref{aweights}.


\section*{Acknowledgement}
We would like to thank Professor Richard J. Gardner from Western Washington University in the USA and Professor Peter Gritzmann from Technical University of Munich in Germany for helping us making the connection between a convex hull and LP-solve.

\end{document}